\documentclass[sigconf]{acmart}

\usepackage[english]{babel}

\usepackage{algorithm}
\usepackage[noend]{algpseudocode}
\usepackage{subfig}
\usepackage{xcolor}

\usepackage{booktabs}
\usepackage{outlines}
\usepackage{color}
\usepackage{amssymb}

\usepackage{multirow}

\hypersetup{draft}

\newcommand{\eg}{{\it e.g.}}
\newcommand{\ie}{{\it i.e.}}
\newcommand{\etc}{{\it etc.}}

\definecolor{amber}{rgb}{1.0, 0.49, 0.0}
\definecolor{ao}{rgb}{0.0, 0.0, 1.0}
\definecolor{fashionfuchsia}{rgb}{0.96, 0.0, 0.63}
\definecolor{ballblue}{rgb}{0.13, 0.67, 0.8}
\definecolor{darksalmon}{rgb}{0.91, 0.59, 0.48}
\definecolor{ferngreen}{rgb}{0.31, 0.47, 0.26}
\definecolor{blue-violet}{rgb}{0.54, 0.17, 0.89}
\definecolor{chamoisee}{rgb}{0.63, 0.47, 0.35}

\setcopyright{none}
\acmConference{ACM}{}

\renewcommand\footnotetextcopyrightpermission[1]{} 

\begin{document}

\title{Beyond content analysis: Detecting targeted ads via distributed counting}

\author{Costas Iordanou}
\affiliation{
	\institution{MPI, Germany}
}
\email{iordanou@mpi-inf.mpg.de}

\author{Nicolas Kourtellis}
\affiliation{
	\institution{Telefonica Research, Spain}
}
\email{nicolas.kourtellis@telefonica.com}

\author{Juan Miguel Carrascosa}
\affiliation{
	\institution{Cyprus University of Technology}
}
\email{mcarrascosa@gmail.es}

\author{Claudio Soriente}
\affiliation{
	\institution{NEC Laboratories Europe}
}
\email{claudio.soriente@emea.nec.com}

\author{Ruben Cuevas}
\affiliation{
	\institution{Universidad Carlos III de Madrid (Telematics Engineering Department \& UC3M-Santander Big Data Institute)}
}
\email{rcuevas@inv.it.uc3m.es}

\author{Nikolaos Laoutaris}
\affiliation{
	\institution{IMDEA Networks Institute}
}
\email{nikolaos.laoutaris@imdea.org}

\renewcommand{\shortauthors}{Iordanou et. al.}

\begin{abstract}
Being able to check whether an online advertisement has been targeted is essential for resolving privacy controversies and implementing in practice data protection regulations like GDPR, CCPA, and COPPA. In this paper we describe the design, implementation, and deployment of an advertisement auditing system called \textit{eyeWnder} that uses crowdsourcing to reveal in real time whether a display advertisement has been targeted or not. Crowdsourcing simplifies the detection of targeted advertising, but requires reporting to a central repository the impressions seen by different users, thereby jeopardizing their privacy. We break this deadlock with a privacy preserving data sharing protocol that allows \textit{eyeWnder} to compute global statistics required to detect targeting, while keeping the advertisements seen by individual users and their browsing history private. We conduct a simulation study to explore the effect of different parameters and a live validation to demonstrate the accuracy of our approach. Unlike previous solutions, \textit{eyeWnder} can even detect indirect targeting, \ie, marketing campaigns that promote a product or service whose description bears no semantic overlap with its targeted audience.
\end{abstract}

\maketitle

\section{Introduction}
\label{sect:introduction}

Targeted advertising offers the possibility of delivering tailored advertisements (in the following \emph{ads}) to users based on their interests and demographic properties.
It has helped the advertising industry reach high growth rates and revenues (23.2\% growth and \$23.9B in Q1 '18 in US only)~\cite{iab2018_q1}, has created lots of jobs, subsidized the delivery of free services, and funded a lot of digital innovation. Of course, to deliver targeted ads, so-called AdTech companies need to detect users' interests and intentions which is done by monitoring visited pages,  searched terms, social network activity, \etc

In its principle, the concept of targeted (or personalized) advertising appears benign: offering to consumers products and services that they truly care about, instead of irrelevant ones that distract or annoy. It is in its implementation and actual use where controversies start arising. For example, tracking should respect fundamental data protection rights of people, such as their desire to opt-out, and should keep clear from sensitive personal-data categories, such as health, political beliefs, religion or sexual orientation, protected by data protection laws like GDPR~\cite{EU-GDPR} in Europe and the California Consumers Privacy Act (CCPA)~\cite{CCPA} in US. Similarly, sensitive demographic groups, like children, should be protected from data collection and targeting as mandated, for example, by FTC's COPPA~\cite{coppa} regulation in US. Unfortunately, this is not always the case, as made evident by the continuous presence of the topic in the news and public debates~\cite{bnd_2017, mediainC2017}, or the conducted investigations and placed fines~\cite{AG_Schneiderman}.
A direct consequence of concern around privacy is the rise in popularity of anti-tracking and ad-blocking tools~\cite{adblockplus, adguard, ghostery, noscript, privacybadger}. This surge of software can choke the web of its advertising revenues. To avoid a Tragedy of the Commons triggered by eroding privacy~\cite{Laoutaris18editorial}, companies offering web services need to gain back the trust of their users. An important step in this direction would be to provide users with the ability to single out and report ads that violate privacy norms and laws.

The networking, measurements, and distributed systems community has been active in the development of a new breed of  transparency tools~\cite{Laoutaris18editorial} for end users, data protection authorities, and the advertising sector's own self-regulation initiatives. Initial efforts went to detecting online price discrimination~\cite{Mikians_2012,Mikians2013,Hannak_2014,Iordanou2017} followed by tools for detecting targeted advertising~\cite{Carrascosa_2015, xray, Sunlight, MyAdChoices_Jagdish16, AdReveal, Venkatadri_2018}. Having the ability to verify the deja vu feeling arising when running into seemingly familiar offerings is important for user empowerment and goes beyond mere curiosity. For example, it can allow a user to check whether marketers have respected his opt-out signals, expressed through Do-Not-Track~\cite{dnt}, AdChoices~\cite{adChoice}, or any of the numerous self-regulation initiatives put forth by the advertising sector. Moreover, if a user runs into an ad related to sensitive, and hence protected, data category, he should be able to verify whether this has been targeted, or is appearing for other reasons. The latter is not always trivial, as with the so called "re-targeted" ads displaying products and services visited by the user in the recent past. Behavioral targeting can be more general and, for example, display to a user a pair of sneakers of brand B' at store S' because the user viewed a pair of sneakers of brand B at store S. A pair of jeans may then be displayed due to the previously viewed sneakers. Detecting such indirect targeting quickly becomes a guessing game, and is, therefore, of little use to the monitoring and enforcement of data protection laws.

Early work on detecting targeting has employed artificially created ``personas'', \ie, browsers scripted to visit certain pages that allude to clear demographic types~\cite{Carrascosa_2015, xray, Sunlight, Bashir_2016}, as ``bait'' for measuring whether ad delivery channels target these demographic types. In the ``offline'' version of the same approach, researchers have looked at passively collected web click-streams to detect correlations between the pages visited by real users and the ads delivered to them~\cite{Wills_2012, Adscape:2014, AdReveal}. While these studies have made important contributions, in general, they have been designed to operate offline, at a low scale, or using simulated personas. In this paper, we aim at addressing the following question: ``\emph{Can we detect ad targeting with real users, in real time, and at large scale?}''

\noindent \textbf{Our contributions:} In this paper, we propose a novel, scalable, and real-time ad detection approach, and implement it on real user's devices.
Instead of using ``personas'' and automated bots to collect ads from artificial visits to pages, we rely on a custom protocol for collecting statistics about the actual ads encountered by real users while browsing online. We show that a surprisingly simple count-based heuristic can detect targeting with high precision. This simple heuristic is based on the observation that targeted ads tend to ``follow'' specific users across multiple domains, while being seen by relatively fewer users than non-targeted ones. The heuristic can be computed in real-time and in a scalable fashion. Furthermore, it is agnostic to how users' information was collected, and to how impressions were auctioned and delivered. This ``black box'' approach only looks for correlations between users and advertisements, and is, thus, robust to detection countermeasures~\cite{bashir-websockets-imc18}. Evading it, would require eliminating such correlations, which goes against the spirit and the essence of targeted advertising.

Crowdsourcing is a powerful tool for detecting targeted ads, but requires users to report the ads they encounter in different websites. The second technical contribution of our work is a protocol for exchanging this information in a privacy-preserving manner. We leverage a wealth of previous work on privacy-preserving aggregate statistic computation~\cite{Castelluccia09,MelisDC16,KursaweDK11}, to compute aggregate statistics required by the ad detection algorithm, while keeping the ads seen by users\footnote{Note that targeted ads can reveal user interests~\cite{Castelluccia:2012}.} and their browsing history private. In particular, we design a privacy-preserving protocol to compute distribution of ads seen by users. Different from previous work in this area, we face the challenge of protecting not only the distribution of the ads, but the ads themselves. This problem was recently framed by~\cite{fanti16popets} as the problem of estimating ``unknown unknowns''. The techniques presented in~\cite{fanti16popets} are based on differential privacy and require clients to report their ``real'' distribution as well as distributions computed using n-grams of the labels (in our case, ad URLs). As such, their technique fits scenarios where labels are short and not random. We take a different approach and propose a technique that is less involved and works with any label size.

We have implemented our count-based heuristic and the privacy-preserving protocol to support it in a distributed system that we call \textit{eyeWnder}. The third contribution of our work is the deployment, validation, and measurement study executed using \textit{eyeWnder}.

\noindent \textbf{Our findings:} We have been operating \textit{eyeWnder} live for more than one year with close to 1000 users in order to get feedback on matters of user experience and desired features. This user base includes 100 paid volunteers from FigureEight~\cite{FigureEight} who have agreed to lend their data to our validation efforts. We have also conducted extensive simulation studies in which we could control how advertisements are displayed. Our findings are as follows:

\noindent - Our simulations show that it only takes 6 to 7 repetitions of an ad to make it detectable by \textit{eyeWnder}. Even with such low repetition frequency, our algorithm achieves a false negative rate of less than 30\%. Tuning the algorithm can bring the false negative rate to below 10\%, at the expense of requiring around 5 more repetitions.  More importantly, false positives, i.e., non-targeted ads classified as targeted, are typically close to zero, and reach up to 2\% only in the most extreme corner scenario that we have evaluated. Having very low false positives means that when \textit{eyeWnder} classifies an ad as targeted, then it most probably is. This is fundamental, if the tool is to be used for reporting suspected data protection violations.

\noindent - The above simulations are aligned with the results from our live validation with real users and advertising campaigns that are not under our control. Count-based ad detection, even with as few as 100 users, yields high precision, with true positive and true negative rates of 78\% and 87\%, respectively.

\noindent - Privacy-preserving crowdsourcing allows to compute the aggregate statistics required by the ad detection mechanism while keeping user browsing history and received ads private. In particular, the privacy-preserving protocol has a negligible effect on the quality of the computed statistics.

\noindent - We have detected several examples of indirect targeting that existing content-based techniques (\ie, semantic overlap between the user profile and the received ad) are unable to detect, as well as signs of advertising bias towards different demographic traits. 

\noindent - We make available our plugin via the Chrome Store, for wide testing and use from end-users, privacy researchers and auditors\footnote{\url{http://www.eyewnder.com/views/index}}.

\section{Background and Requirements}
\label{sect:background}

\subsection{Types of online ads}
\label{subsect:typesoftargeting}

In the rest of the paper, we will refer to ads as either targeted or non-targeted.
Furthermore, we will distinguish targeted ads between Directly- and Indirectly-targeted ads.

\noindent \textbf{Targeted vs. Non-targeted ads:}
\noindent \emph{Targeted ads} are selected based on data about the user visiting a website. The exact algorithms used are trade secrets of the different AdTech companies but it is widely accepted that such algorithms rely upon user demographic information (gender and age), geo-location information (GPS coordinates, IP address or Base-station id), behavioral information extracted from user online activity (websites visited, searched terms, social network activity, \etc).  The form of targeted advertising based on behavioral attributes is referred to as \emph{Online Behavioral Advertising (OBA)} whereas ads regarding a previously visited webpage offering products or services (\eg, a website offering hotels) is referred to as \emph{Retargeting}~\cite{Bashir_2016}.

\noindent \emph{Non-Targeted ads} are shown irrespectively of the user visiting the website. This class of ads includes static ads (shown to all users visiting a website based on a private deal between advertisers and publishers), as well as contextual ads (ads matching the context/topic of the website, \eg, a sports ad appearing on a sports website).

\noindent \textbf{Direct vs. Indirect targeting:}
\noindent \emph{Direct targeting} is the most obvious form of targeting. In this case, an advertiser interested in selling products of a certain category (\eg, fishing products) targets users tagged as interested in such category (\ie, fishing). Most of previous work has focused on analyzing this type of targeting~\cite{Carrascosa_2015, xray, Sunlight}.

\noindent \emph{Indirect targeting} is applied when marketers may target a certain group of users with offerings that have no direct semantic overlap with the category this group has been tagged with. For instance, it has been reported that fans of the Walking Dead TV series were targeted with pro Donald Trump material~\cite{WalkingDead}.
This would be an example of indirect targeting, since the targeted user group (Walking Dead fans) has no immediate semantic overlap with the advertised offering.
To the best of our knowledge, \textit{eyeWnder} is the first proposal that tackles indirectly targeted ads.

\subsection{Requirements}
\label{subsect:requirementsForOBA}

Next, we enumerate the most important requirements of a system for real-time detection of targeted ads. Such requirements have emerged from reviewing the limitation of existing approaches found in the literature. In Section~\ref{sect:related_work}
and Table~\ref{table:comparison_rw} we provide an extensive comparison between our solution and existing proposals.

\noindent \textbf{Generality:}
The detection mechanism should be able to analyze any web-based display ad. It should not be limited to a specific ecosystem (\eg, Facebook advertising) where more information about the user or the ad may be available, or platform (\eg, AdChoices initiative~\cite{adChoice}).
Moreover, detection should work independently of the tracking mechanisms used for collecting user data, interests, and intentions (be it via cookies, browser fingerprinting, or user contributed information such as searches or social network activity).

\noindent \textbf{Precision:}
The mechanism should allow untrained users for auditing, and potentially reporting, offending ads. Therefore it is essential to have a high detection precision in terms of True Positives (TP) and True Negatives (TN). Validation should be carried out using publicly available data, \ie, without requiring special access to silo-ed data from AdTech delivery channels, Telcos, marketing campaigns, or other gatekeepers of such information that, generally speaking, have no incentive to make it public.

\noindent \textbf{Simplicity:}
The detection method should be as simple as possible, so that it can be implemented as a distributed system with most of the functionality and code running on the end-user device (browser).

\noindent \textbf{Real time operation:}
A user should be able to request auditing of a particular ad appearing in his browser, and the system should respond within at most few seconds.

\noindent \textbf{Scalability:}
The detection method should be able to handle a large number of users (in the order of tens of thousands) without special requirements on the back-end, including CPU load, memory, storage, and bandwidth consumption. Resource consumption must be limited to ``control plane'' rather than ``data plane'' tasks, to make the method scale.

\noindent \textbf{Ad-fraud avoidance:}
Differently from previous work~\cite{Carrascosa_2015, xray, Sunlight}, the method should avoid fake visits to pages, since fake visits contribute to fake ad impressions and click-fraud.

\noindent \textbf{Detection of indirect targeting:}
The detection method should be able to detect both direct and indirect targeting.

\noindent \textbf{User privacy protection:}
The method should not jeopardize the privacy of end-users by, \eg, requiring them to share sensitive data such as their browsing history or ads seen.

\section{Ethical considerations}
\label{sect:ethical}

We have obtained ethical approval from our institutions and the funding agencies to conduct this research. 
Moreover, we have obtained explicit consent from users, before installing our extension through the FigureEight platform interface~\cite{FigureEight}, to collect and process the anonymous data used in this paper.

\section{A Count-based algorithm}
\label{sect:countbasedalgo}

In this section, we describe a count-based algorithm for detecting targeted ads using only frequency counts of impressions seen by users across different domains. The algorithm is inspired by simple observations on how targeted ads behave, namely 1) targeted ads tend to ``follow'' targeted users across multiple domains, and 2) targeted ads are seen by relatively fewer users than non-targeted ads.

\subsection{Algorithm description}
\label{subsect:countbaseDescription}

Our algorithm is simple and is based on the above remarks.
By observation (1), if a given ad $\alpha$ is targeting a specific user $u$, that user is likely to encounter $\alpha$ across multiple domains. Therefore, the algorithm counts the number of different domains ($\text{\#Domains}_{u,\alpha}$) where user $u$ has seen $\alpha$, and labels the ad as targeted if that number crosses a threshold that we call $\text{Domains}_{{\text{th},u}}$.
Similarly, by observation (2), if $\alpha$ is targeting $u$, most likely very few other users will see $\alpha$ during their browsing activity, i.e., only users that share similar interests with $u$. Therefore, the algorithm counts the number of different users ($\text{\#Users}_{\alpha}$) that have seen $\alpha$ and labels $\alpha$ as targeted if that number is below the $\text{Users}_{\text{th}}$ threshold. Note that the algorithm annotates an ad as targeted if both conditions holds.

Given an ad $\alpha$, the number of domains where a user has seen $\alpha$, along with the corresponding $\text{Domains}_{{\text{th},u}}$ threshold are dependent on user $u$ and, thus, can be computed locally. On the other hand, computing the number of different users that have seen $\alpha$, as well as the $\text{Users}_{\text{th}}$ threshold requires a global view of the system.

\subsection{Algorithm details}
\label{subsect:countbaseInternals}

\noindent\textbf{Threshold estimation:}

A fundamental design choice of our algorithm is how to compute the $\text{Domains}_{{\text{th},u}}$ and $\text{Users}_{\text{th}}$ thresholds. We empirically evaluated different options based on several moments of the distributions (the mean, the median, the standard deviation, and possible combinations thereof). We eventually settled for the mean of the distribution since it offered the best trade-off between accuracy and the data we require from our users.
For the sake of clarity, we do discuss all the alternatives we have considered but, in Section~\ref{subsec:sim_results}, we demonstrate the performance difference for a few of these options and show why we settle for the mean.

In order to be able to set the $\text{Domains}_{\text{th},u}$ threshold, we require a minimum amount of information from users. Similar to~\cite{bernard1986densityestimation}, we require that users have visited at least 4 domains that serve ads within the last 7 days. If this minimum requirement is not met, our algorithm refrains from making a guess for lack of sufficient data.

\noindent\textbf{Time-window selection:} Our algorithm operates in time intervals in order to update the $\text{Users}_{\text{th}}$ threshold, since users can continuously receive and report ads during their normal browsing sessions. We select a time window of one week based on the following observations. 
First, users tend to browse differently during weekdays and weekends~\cite{Boughareb11weeklybrowsing}. Therefore, we consider a week as a natural time period where both weekdays and weekend are captured, allowing us to study users' behavior in both types of days. Second, we consider how targeted ads behave, which aggressively follow the user for a few days and gradually fade-out over time. Thus, a window of seven days is sufficiently large to capture targeting and allows to collect enough historical information for the algorithm to operate. In fact, we directly contacted 4 large DSP platforms and confirmed that the majority of ad-campaigns they serve last a week or more. Note that the $\text{Domains}_{\text{th},u}$ threshold is per individual user, thus, can be updated in real-time within each user's browser.

\section{The \textit{eyeWnder} system}
\label{sect:eyeWnder}

In this section, we describe the high level components of the system and the information flows between them.

\begin{figure}[t!]
  \begin{center}
    \includegraphics[scale=0.64]{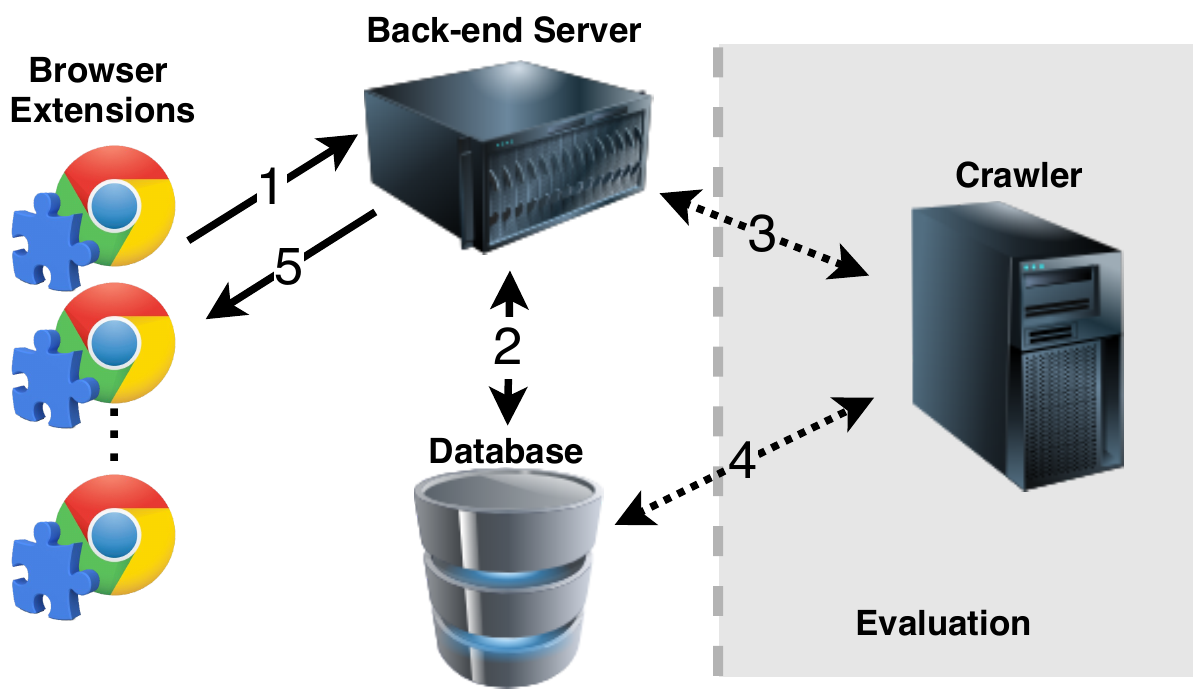}
    \vspace{-3mm}
    \caption{Architecture overview of the \textit{eyeWnder} system and the information flow between entities.}
    \label{fig:eyeWnderDiagram}
  \end{center}
  \vspace{-4mm}
\end{figure}

The high level architecture of \textit{eyeWnder} is depicted in Figure~\ref{fig:eyeWnderDiagram}.
The system consists of four components: the browser extension instances, a back-end server, a centralized database, and a crawler server.

\noindent\textbf{Browser extension:} The extension performs the following functions:
(1) Collects information about the ads rendered to the user.
(2) Reports information to the back-end server (Figure~\ref{fig:eyeWnderDiagram}, arrow 1) through the privacy-preserving protocol of Section~\ref{sect:privacy}.
(3) Classifies ads as targeted or non-targeted, by leveraging the algorithm of Section~\ref{subsect:countbaseDescription}.

The actual algorithm that classifies ads as targeted vs. non-targeted consists of just a few lines of JavaScript code. Identifying and collecting info about ads is more involved. Indeed, different ad delivery channels use a multitude of techniques for delivering display ads.
Several of them go at great lengths to make programmatic detection of their code difficult, in an attempt to evade ad-blocking software. Our extension runs an ad-detection algorithm to automatically detect ads within a page, and a landing page detection algorithm to infer the pages where ads lead once clicked.
Our ad-detection algorithm is similar to the one of AdBlockPlus~\cite{adblockplus}. Our goal, however, is just to analyze an ad, and not to block it.

In order to avoid click-fraud, the landing page detection algorithm does not click on the ad, but rather applies a series of heuristics to discover the landing URL.
In particular, the algorithm examines \texttt{<a>} HTML tags to extract the URL from the \texttt{href} field; alternatively, the algorithm looks for \texttt{onclick} events and extracts the URL if it exists\footnote{In some cases the \texttt{onclick} event is redirected to a JavaScript function instead of a URL redirection.}.
In case of JavaScript code, we run a regex to detect any URL-like strings within the script text.
In all the above cases, if the detection algorithm finds a URL that does not belong to well-known ad networks, we consider the URL as the ad landing page. Otherwise, we refrain from resolving the URL in order to avoid click-fraud.  A similar technique was used in ~\cite{Javaid_2017,MyAdChoices_Jagdish16}. In case of ads with randomized landing page URLs (e.g., malicious ads~\cite{Thomas_2015} or customized dynamic ads~\cite{Plp_patent}), we use the ad content (\ie, the image URL, \etc) to uniquely identify the same advertisement across different impressions. To identify ad networks that use randomized landing page URLs we use the methodology described in \cite{Butkiewicz_2015}.

\noindent\textbf{Back-end server:}
The server maintains the $\text{\#Users}_\alpha$ counters and computes the corresponding $\text{Users}_{\text{th}}$ threshold.
By leveraging the privacy-preserving protocol of Section~\ref{sect:privacy}, the server receives \emph{blinded} reports from the extensions, aggregates them, and extracts an estimate of the number of users that have seen an ad---thereby computing the $\text{\#Users}_\alpha$ counters. The server also computes the $\text{Users}_{\text{th}}$ threshold by applying the methodology of Section~\ref{subsect:countbaseInternals}; the computed threshold is then distributed to clients (Figure~\ref{fig:eyeWnderDiagram}, arrow 5).

\noindent\textbf{Database:}
We use MySQL to store system metadata, such as, active users within the system, historic anonymized data reported by users, \etc (Figure~\ref{fig:eyeWnderDiagram}, arrow 2). We also store aggregated data that we need for evaluation purposes.

\noindent\textbf{Crawler server:}
This component is used only for evaluation purposes, and is responsible for collecting ad-related data on specific webpages. The crawler is controlled by the back-end server and can be instructed to visit specific websites upon request (Figure~\ref{fig:eyeWnderDiagram}, arrow 3). Specifically, the crawler server visits audited pages to collect ads with a clear browsing profile (empty browser cache and an empty set of cookies). These ads are then used for deciding whether \textit{eyeWnder} has indeed classified accurately an ad as targeted (in which case the crawler should not encounter it during a visit).
The crawler can launch multiple instances of a clean profile browser with the \textit{eyeWnder} extension installed, and store the detected ads directly into the database (Figure~\ref{fig:eyeWnderDiagram}, arrow 4).

\section{Privacy Preserving Protocol}
\label{sect:privacy}

In this section, we detail a privacy-preserving protocol that allows the back-end server to compute the $\text{\#Users}_\alpha$ counters while clients keep the ads they have seen private.

We leverage techniques used in many proposals for privacy-preserving aggregated statistics~\cite{Castelluccia09,MelisDC16,KursaweDK11}.
The basic idea is that each user \emph{blinds} his report before sending it to the server.
Blinding factors are agreed upon by all users and are such that if the server aggregates all reports, the blinding cancels out and the aggregate statistic (\eg, the sum of all reports) becomes available to the server.
In a nutshell, one can think of the blinding factors as additive random shares of $0$.
If users report a vector of values, they should compute separate blindings for each position of the vector.
One fundamental assumption underlying the design just described is that all parties in the system (\textit{i.e.}, users and the server) can enumerate the whole set of elements to be reported.
In our scenario, this set---we denote it by $A$---includes all ads seen by at least one of our users. Therefore, its size may be large and, most importantly, users may not be able to enumerate it. For example, user Alice may not know what ads have been seen by user Bob. In such settings, one could use synopsis data structures for multi-sets that admit aggregation. For example count-min-sketches~\cite{CMS05} (CMS) or spectral bloom filters~\cite{SBF_Cohen03} can be used.
In this work, we use CMS as they allow us to bound the probability of error, as well as the error itself.

\subsection{Count-min-sketch}

A CMS X is a bi-dimensional array with $d=\lceil \ln{T/\delta}\rceil$ rows and $w=\lceil e/\epsilon\rceil$ columns, where $T$ is the number of elements to be counted.
All the cells of the sketch are initialized to 0 and $d$ pairwise-independent hash functions $\{h_j:\{0,1\}^*\rightarrow [w]\}_{1\leq j\leq d}$ are chosen.

The encoding of an element $x_i$ is done by calling \texttt{X.update($x_i$)} that increments $X[j,h_j(x_i)]$ by $1$, for $1\leq j\leq d$.

The estimated frequency $\bar{c}_{x_i}$ of element $x_i$ is retrieved by calling \texttt{X.query($x_i$)} that outputs $\min_j X[j,h_j(x_i)]$ such that:

\begin{enumerate}
\item $c_{x_i}\leq \bar{c}_{x_i}$
\item $\bar{c}_{x_i}\leq c_{x_i}+\epsilon\sum_{j=1..T} c_{x_j}$ with probability $1-\delta$
\end{enumerate}

\noindent where $c_{x_i}$ is the true frequency of element $x_i$.

In \textit{eyeWnder}, each user encodes the set of ads he has seen in a CMS data structure, and blinds each cell before sending it to the server. The server aggregates all CMSes so that all blindings cancel out, and obtains the aggregate CMS encoding the multi-set of ads seen across all users.
If users share an additive random share of $0$ for each cell of the CMS, this design allows for a privacy-preserving aggregate statistics framework for scenarios where the set of elements to be counted is (a) large and (b) not enumerable by each party.

This design is similar to the one shown in~\cite{MelisDC16}, that in turn, extends techniques presented in~\cite{KursaweDK11}. However, the clients in~\cite{MelisDC16} can enumerate the set of elements to be reported and, therefore, \cite{MelisDC16} presents a less challenging scenario.

We also face an additional challenge, that, to the best of our knowledge, has not been addressed by existing privacy-preserving aggregate statistic techniques. Data structures like spectral bloom filters or CMS allow to query for the estimate frequency of a given element $x$. That is, the querier (\ie, our server) must enumerate all the ads encoded in the aggregate CMS, in order to learn the distribution of ads as seen by users. Of course, the server cannot do so and our privacy goal forbids clients from sending the URLs of the ads to the server. We overcome this problem as follows.

We map the URL of an \emph{ad  ID} in $[1,|A|]$ by means of a pseudo-random function (PRF). The latter is keyed to prevent the server from computing the mapping on its own. In particular, given an ad URL $x$, its ad ID is computed as $y=F(k,x)$ where $F$ is a PRF and $k$ is a cryptographic key. For each ad seen by a user, the extension computes the corresponding ad ID and encodes it in the CMS. Note that without knowledge of $k$, it is not possible to relate ad URL $x$ to its identifier $y$.

Rather than hard-coding the key $k$ in the extension, we borrow from previous research on Oblivious Pseudo-Random Function (OPRF)~\cite{Freedman_2005} and introduce an additional server to help clients mapping ad URLs to ad IDs. The \emph{oprf-server} holds the secret key $k$ and aids clients to compute $y=F(k,x)$ for a given ad URL $x$. As the name suggests, the server is ``oblivious'' to the input of the PRF so that $x$ remains private to the user.\footnote{We note that in order to avoid a single point of failure, mapping function can be distributed to multiple servers by defining $F$ as the XOR of the output of multiple OPRFs, each computed with its own secret key.}
While such a design choice requires an additional server, we note that previous work uses a similar approach in order to improve the overall security of an application~\cite{pythia,dupless}. In a real-world deployment, the oprf-server may be instantiated by already-deployed trusted-third parties such as certification authorities, EFF, \etc

In practice, we have to (over)estimate $|A|$ in order to minimize collisions when mapping an ad URL to an ad ID. However, by overestimating $|A|$, the server is likely to query the CMS for ad IDs that correspond to none of the encoded ads (\ie, a false positive). Nevertheless, later we show that a CMS is robust to false positives by design.

In the following, we provide details of the protocol.

\noindent\textbf{Blinding factors:} We borrow from Kursawe et al.~\cite{KursaweDK11} who have shown how a set of users can agree on random shares of $0$. In particular, let $N$ be the number of users and $M$ be the number of elements to be blinded (\eg, the number of cells in a CMS).
Also, denote by $x_i,\ y=g^{x_i}$, the private and public key of user $u_i$, respectively.
Here, $g$ is a generator of a cyclic group $G$ of order $q$, where Computational Diffie Hellman is hard. Assume that the public key of each user is available to all other users in the system, \eg, by means of a public bulletin board like an online forum\footnote{The board may be as well hosted at the back-end server.}. At round $s$, user $u_i$ generates a blinding factor for the $m$-th cell as:
$$
b_i[m]=\sum_{j=1,j\neq i}^{N} H(y^{x_i}_j || m || s) \cdot (-1)^{i>j}
$$
where $(-1)^{i>j}$ is equal to $1$ if $i>j$, or to $-1$ otherwise.
Note that each user can locally compute his blinding factors by simply using the public keys of all other users. Note also that for any $m$, we have $\sum_{i=1}^{N} b_i[m]=0$, \ie, $b_i[m]$ is an additive random share of $0$.

\noindent\textbf{OPRF:}
We leverage the RSA-based OPRF in~\cite{Jarecki_2009}. Given an RSA triple $(N,d,e)$ where N is the product of two distinct primes of sufficient length and $d,e\in Z^*_\phi(N)$ are such that $ed\equiv 1\mod \phi(N)$, the PRF on input $x$ is defined as  $G(H(x)^d)$ where $H:\{0,1\}^*\rightarrow Z_N$ is a hash function mapping arbitrary strings to elements of $Z_N$, and $G:Z_N\rightarrow\{0,1\}^l$ $H:\{0,1\}^*\rightarrow Z_N$ is a hash function mapping elements of $Z_N$ to strings of arbitrary length $l$. The oprf-server generates the RSA triple (via a suitable key-generation algorithm) and publishes $N,e$ while keeps $d$ private. Given an ad URL $x$, the client issues a request as $x'=H(x)r^e$. That is, the client maps $x$ to $Z_N$ and blinds the result by multiplying it with a random group element $r$ raised to the $e$-th power. The server ``signs'' the request by computing $y=(x')^d$. Finally, the server recovers $y'=\frac{y}{r}$ and outputs $y=G(y')$. Note  that $y'=\frac{y}{r}=\frac{(x')^d}{r} =\frac{(H(x)r^e)^d}{r}=\frac{H(x)^d r^{ed}}{r} = H(x)^d$ since $ed=1\mod \phi(N)$. The protocol guarantees that the server learns nothing about $x$ whereas the client learns nothing about $d$, and it is a PRF under the one-more RSA assumption~\cite{Jarecki_2009}.

\noindent\textbf{CMS computation:} User $u_i$ starts with an empty CMS $X_i$. For each newly received ad $x$, the user engages in an OPRF protocol with the oprf-server to obtain the corresponding ad ID $y$, and encodes $y$ in the CMS by calling \texttt{$X_i.update(y)$}.
When asked to report its CMS, the user blinds each cell of $X_i$ with the blindings computed as shown above. That is, the client computes the blinded CMS $\hat{X}_i$ by computing $\hat{X}_i[j]=X_i[j]+b_i[j]$ for $1\leq j\leq m$ where $m$ is the number of cells in the CMS. Finally, the client sends $\hat{X}_i$ to the server.

\noindent\textbf{Aggregation and unblinding:} Finally, the server aggregates all received CMSes by computing $X=\sum_i{\hat{X}_i}$ where the sum is cell-wise and, for each ad ID $i\in\{1,\ldots |A|\}$ it obtains the estimated frequency by querying \texttt{X.query($i$)}.

\noindent\textbf{Fault-tolerance:} Our aggregation technique requires that all users report their blinded CMSes, so that the server can aggregate them and cancel out the blindings. If a user fails to report its CMS, aggregation of the remaining ones at the server results in a CMS with random noise in each of its cells. In order to tolerate missing reports, the server and the clients who have sent their reports must go through an additional round of interaction to ``adjust'' their blindings and cater for the blindings of the non-reporting clients (see ~\cite{MelisDC16}). The protocol takes only two rounds---one where the server reports the list of ``missing'' clients and another round where the clients send their CMSes obfuscated with the updated blinding factors.

\noindent\textbf{Security:} The security of our scheme follows from the security of the protocol to compute blindings proposed in~\cite{KursaweDK11}. We tweak the protocol by using an OPRF to map ad URLs to a set that is enumerable by the server. By the security provisions of the OPRF protocol, given an ad ID $y$, it is impossible to retrieve its original URL without knowledge of the key held by the oprf-server.

We stress that our protocol remains secure against an honest-but-curious server, and this is an assumption common to many other systems that address privacy issues when computing statistics on crowdsourced data~\cite{Castelluccia09,MelisDC16,KursaweDK11}.

\section{Evaluation}
\label{sect:evaluation}

In this section, we first evaluate the overhead of our privacy preserving protocol. Second, we present a controlled simulation study to assess the robustness of our algorithm and the effect of key parameters on its performance. Third, we present a live validation of the entire system with real ads and the 100 real users recruited via FigureEight~\cite{FigureEight}.

\subsection{Performance and overhead of the privacy preserving protocol}
\label{subsect:evaluation_performaceAndOverheadsPrivacy}

\begin{figure}[t!]
  \begin{center}
    \includegraphics[scale=0.65]{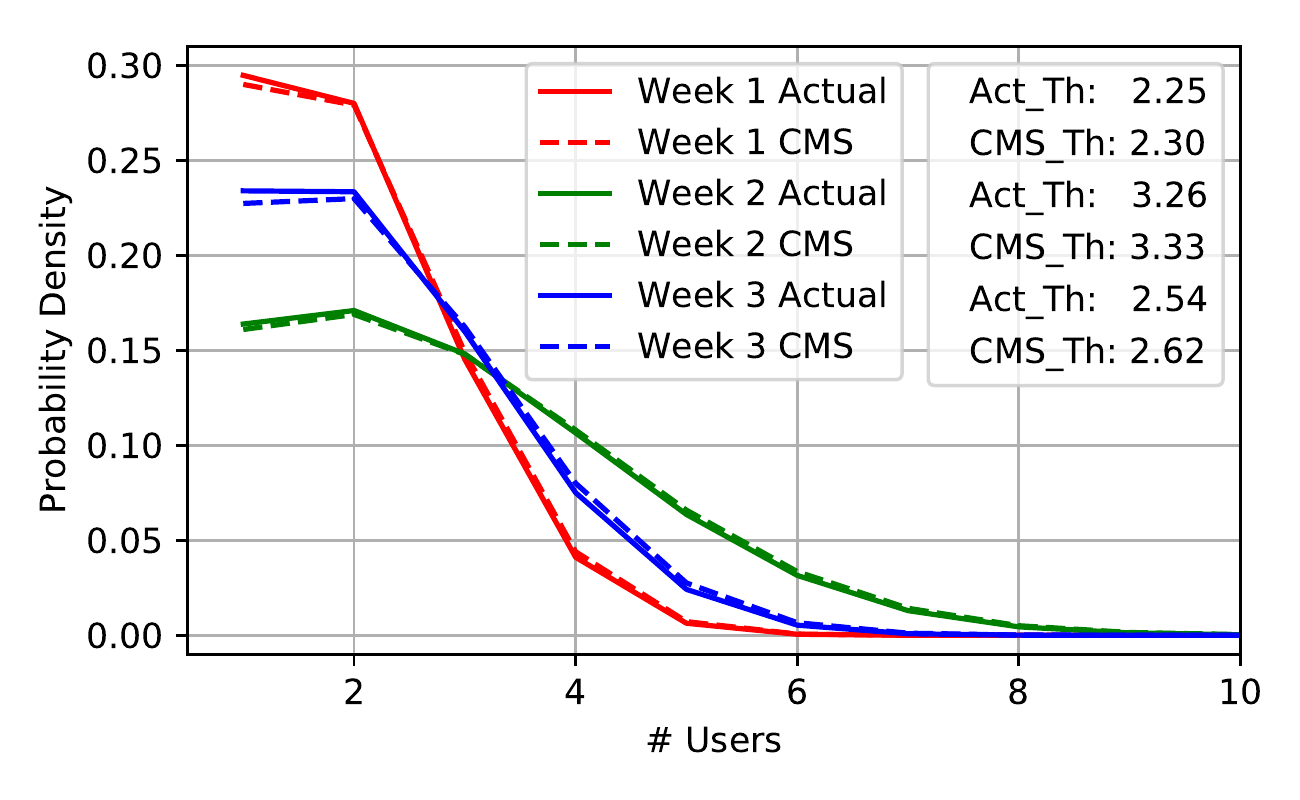}
    \vspace{-8mm}
    \caption{The effect of the privacy preserving protocol on the computation of the \#User distribution and its threshold for three different weeks.}
    \label{fig:actualVsCMS}
  \end{center}
  \vspace{-3mm}
\end{figure}

First, we assess the communication overhead due to the CMS and compare it with the average communication overhead if clients were to upload their contributions in cleartext. We fix $\delta$ and $\epsilon$ to $0.001$ and assume the size of a cell in the CMS to be 4 bytes. The size in bytes of the CMS totals to 185, 196, and 207KB, for an input size\footnote{The number of ads to be counted.} of 10k, 50k, and 100k, respectively. If users were to report their contribution in cleartext, each user would simply report a vector of URLs. Since our dataset suggests that users see 35 unique ads on average, the communication overhead for the average user amounts to roughly 3.5KB (assuming 100-characters URLs using Unicode encoding). Nevertheless, some users with a large number of ads to report can see their communication overhead rise up to hundreds of KB (some users in our dataset reported around 250 unique ads).

Next, we assess the communication and computation overhead required to compute the blinding factors. Exchanged data between the server and a client amount to 0.38MB and 1.9MB for 10k and 50k users, respectively.
The computation time at the client totals 30 seconds for 1k users and a sketch of size ~5k.
Our results are in line with results reported in~\cite{MelisDC16}.
We stress that both operations are carried out once per week and can run in the background.

Also, the time to map the URL of an ad to its ID, using the oprf-server, is always less than $500$ms and requires exchanging two group elements (e.g., 1024 bits each). We stress that the mapping is done once per (unique) ad. It can be carried out as ads are received and results can be stored locally so that they are available when the CMS must be computed.

Finally, we empirically show the effect of the privacy preserving protocol on the computation of the $\text{\#User}_\alpha$ distribution and its threshold. Figure~\ref{fig:actualVsCMS} compares the distribution computed with cleartext reports versus the distribution computed using the privacy-preserving protocol based on (blinded) CMSes. The figure shows the difference using data from three different weeks in our dataset. Furthermore, the figure reports the threshold value computed on cleartext data, and the threshold value computed on the outcome of the privacy-preserving protocol. The latter leads to a slightly higher threshold and this is due to the collisions that may happen when mapping ad URLs to ad identifiers.

To sum up, the overhead due to privacy-preserving protocol does not impose an unbearable toll on users. The protocol to map ad URLs to identifiers is very lightweight and the mapping is carried out as new ads are encountered. The protocol to report the CMS requires a few (\ie 2 or 3) MB of data to be exchanged, assuming 50k users. This is done once per week and the communication complexity scales linearly with the number of users. Further, the error on the estimated distribution due to the privacy-preserving protocol is small as shown in Figure~\ref{fig:actualVsCMS}.

\subsection{Controlled simulation study}
\label{subsect:simulation}

The count-based algorithm of Section~\ref{subsect:countbaseDescription} and its automated parameter tuning described in Section~\ref{subsect:countbaseInternals} can lead to false negatives, \ie, targeted ads classified as non-targeted, and false positives, \ie, non-targeted ones classified as targeted. In this section, we study the circumstances and the frequency of such misclassification via controlled simulation experiments. Out of the two metrics, the most important are the false positives. This has to do with a major use case for \textit{eyeWnder}, which is to report illegal targeting, as explained in the introduction. Therefore, when \textit{eyeWnder} classifies an ad as targeted we want this to be precise with high probability so that investigations are not triggered by mistake. On the other hand, failing to detect a targeted ad is relatively less important since no investigation is launched in this case.

\subsubsection{False negatives} As described earlier, our algorithm uses the fact that the same ad ``follows'' a user across multiple domains as an indication of targeting. But how intense should this following be before our algorithm has a chance to detect it? If a targeted ad appears only once, then certainly it is indistinguishable from any non-targeted ads.
Considering the other extreme, if it appears in all the pages visited by a user, then it obviously becomes easy to detect. Next, we study the effect of the \emph{Frequency Cap} i.e., the number of repetitions (or re-appearances) of a targeted ad on the ability of our algorithm to detect and classify it correctly. Notice that this parameter is not known to us, nor uniform across advertisers and therefore we study its effect in our simulation model by assigning to it different values. The Frequency Cap is used by advertisers to avoid annoying targeted users with too many repetitions of the same ad. In our validation study appearing later in Section~\ref{subsect:accuracy} we demonstrate that our system is indeed robust to the magnitude of values used in practice, but remaining unknown to us.

\subsubsection{False positives} The same ad can be encountered in multiple websites without being targeted. Such is the case, for example, of large-scale ``brand awareness'' campaigns paid by large corporations to display their offering in many mainstream and even niche websites without targeting any particular individual. Such ads will appear to be chasing a user across websites, when in reality they are not. Of course, our algorithm also checks to see that the ad is not seen by many other users of \textit{eyeWnder}, but non uniform user interests can lead to misclassification as shown next.

\subsubsection{Simulation results}
\label{subsec:sim_results}

We have built a custom simulator, based on~\cite{Burklen_2005}, capable of simulating users, websites, and ad campaigns. The main parameters of the simulator and pre-set values for our basic configuration, are depicted in Table~\ref{Table:sim_param}.

\begin{table}[t!]
\caption{Simulation configuration parameters}
\label{Table:sim_param}
\begin{tabular}{lr}
\hline
Variable name              & Value \\ \hline
Number of users            & 500   \\
Number of websites         & 1000  \\
Average user visits        & 138   \\
Average ads per website    & 20    \\
Percentage of targeted ads & 0.1   \\
\hline
\end{tabular}
\end{table}

\begin{figure}[t!] 
  \begin{center}
    \includegraphics[scale=0.67]{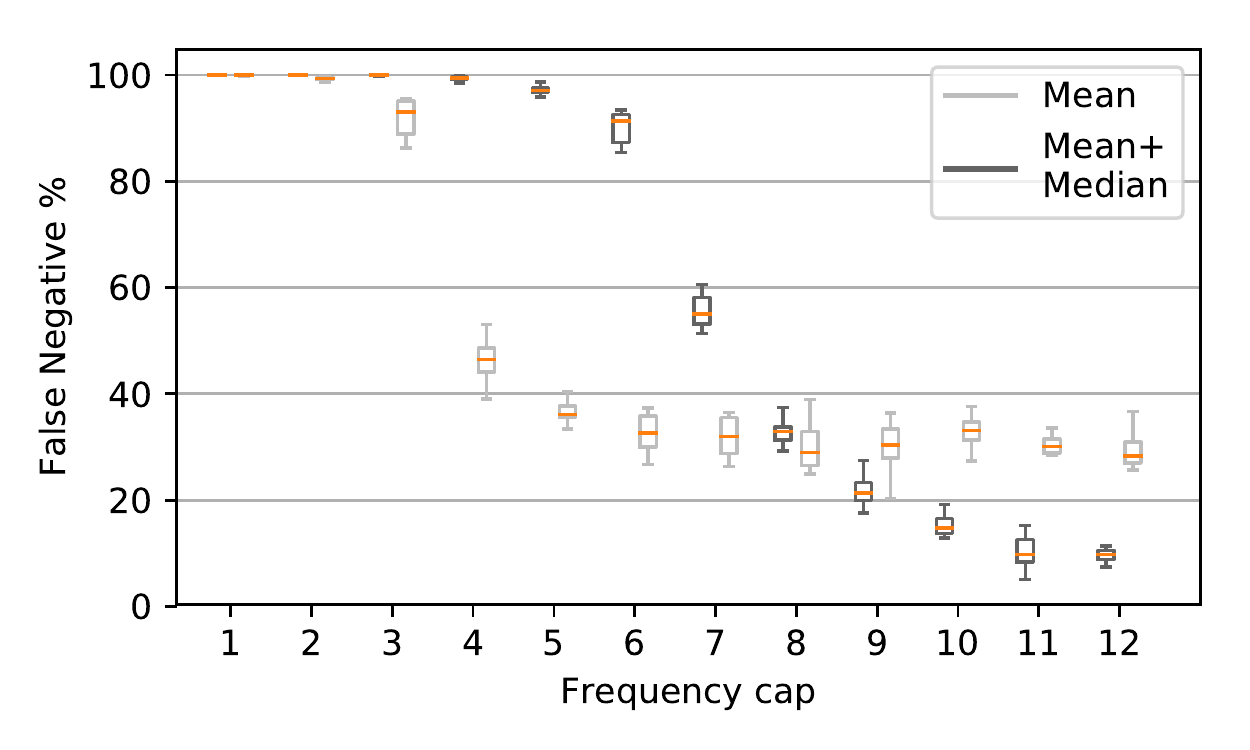}
    \caption{False Negatives \% Vs. Frequency Cap using two different thresholds (Mean, Mean+Median) for both variables ($\text{\#Users}_\alpha$, $\text{\#Domains}_\text{u,a}$)}
    \label{fig:fn_cap}
  \end{center}
\end{figure}

Figure~\ref{fig:fn_cap} shows that even few repetitions of a targeted ad make it detectable by our algorithm with high probability. Setting the $\text{Users}_\text{th}$ and $\text{Domains}_{\text{th},u}$ thresholds according to the \emph{Mean} value of the corresponding counters (see Section~\ref{subsect:countbaseInternals}) brings false negatives to below 30\% with just 6-7 repetitions of an ad. Using as threshold the \emph{Mean+Median} value requires a higher number of repetitions for detection but drops false negative even further to 10\%. Our experiments with real users in the next section are in agreement with these results. In those experiments we set the thresholds using Mean to make sure we can have detection even with a smaller number of repetitions for a targeted ad.

Regarding false positives, we have run several different simulations in which a subset of users visits a subset of sites that happen to be running large static campaigns. These users get to see the same ad in different websites not because they are being targeted but because the ad happens to be in all the websites they visit. If the remaining users (the majority) visits different websites, then misclassification may appear for the initial subset of users. Still, this happens with probability below 2\% in more than 30 different  parameter configurations that we have tried. This result is also validated by our experiments with real users. False positives can be further reduced by grouping users in more homogeneous groups in terms of browsing patterns (e.g., geographically or based on age group, \etc).

\subsection{Live validation of \textit{eyeWnder}}
\label{subsect:accuracy}

\begin{figure*}[t!]
  \begin{center}
    \includegraphics[scale=0.65]{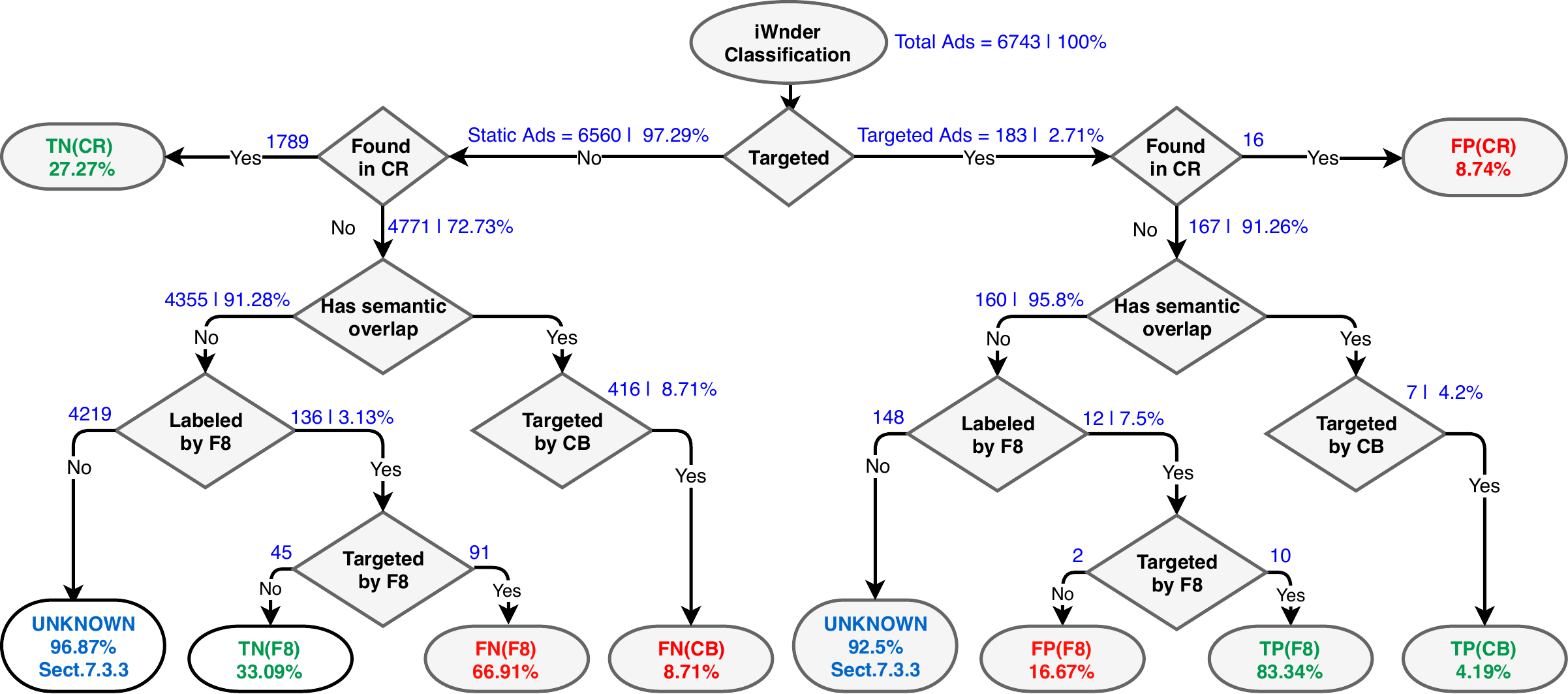}
    \caption{The evaluation tree for precision performance of the \textit{eyeWnder} classification of targeted ads.}
    \label{fig:evaluationTree}
  \end{center}
\end{figure*}

Our simulation results have shown that when \textit{eyeWnder} classifies an ad as targeted then this is accurate with very high probability (false positives <2\%), whereas when it classifies an ad as non-targeted the accuracy is again high (false negatives 10-20\%). In this section we want to validate the above results with a live experiment involving real users and campaigns. This is inherently difficult due to the lack of publicly available ground truth about which ads are targeted and which not (see Requirements in Section~\ref{subsect:requirementsForOBA}). The situation becomes even more challenging when the evaluation has to be performed for a system used by real end-users (as opposed to offline evaluation based on network traces), over a variety of ad delivery channels (as opposed to a single channel, \eg, facebook ads~\cite{Cuevas_2019, Ribeiro_2019, Venkatadri_2018}). Still in this section we perform such a live validation and show that it leads to consistent results with those of our simulation study.

\subsubsection{Datasets}
\label{subsect:evaluation_dataset}

We use three different datasets. The first dataset is created with the help of a crowdsourcing platform named FigureEight~\cite{FigureEight}. We collected data during three consecutive weeks from a population of 100 users with varying level of activity. The dataset includes in total 6743 ads. We call this dataset the \textit{``eyeWnder dataset''} and we use it as input to the count-based algorithm that classifies the collected ads. In addition, we also ask the FigureEight users to label the ads that they receive as targeted or not. For this purpose, we instruct the users to provide labels to all the ads they receive while surfing the web. We call this subset of labeled ads the \textit{``F8 dataset''}.

The last dataset we use is collected by the crawler during the same three weeks.
It includes several visits and ad collections to any website in which \textit{eyeWnder} has classified an ad\footnote{Note that for evaluation we are using full information on our test users after having been granted full consent. The privacy preserving protocol is developed for the actual operation of the system beyond evaluation.}.
We call this dataset, \textit{``CR dataset''} and use it to identify statically placed ads as done in~\cite{Carrascosa_2015, MyAdChoices_Jagdish16}.

\subsubsection{Methodology}

Figure~\ref{fig:evaluationTree} presents the work-flow of our evaluation methodology. The idea is to compare the classification derived by \textit{eyeWnder} against the classification from the crawled dataset (denoted by CR in the figure), the classification from a content-based detection heuristic (CB)\footnote{We have adapted the methodology of~\cite{Carrascosa_2015} to operate with real users instead of personas. For each user, we have selected the most significant categories of the pages he visits to create his profile. We used categories appearing at least T times in different websites. We have used T = 20 since in the context of this paper we are seeking precision rather than recall. We classify an ad as targeted if the main category of its landing page (as obtained from AdWords) matches one of the categories in the user profile.}, and the classification from FigureEight users (F8). These comparisons yield False Positives (FP), False Negatives (FN), True Positives (TP), True Negatives (TN), as well as UNKNOWN rates.
As we will explain shortly, results derived by comparing against the crawler are \emph{correct with high probability} (namely FP(CR), TN(CR)). Results derived by comparing against the content-based heuristic or FigureEight are \emph{likely correct} (namely TP(CB), FN(CB), TP(F8), TN(F8), FP(F8), FN(F8)). The justification for the above is that the content-based heuristic and FigureEight are themselves subjective means of classifying an ad. Users have limitations in detecting bias or discrimination~\cite{Plane:UserPerceptions17}, whereas the content-based heuristic is just another heuristic whose validation faces the same challenges with the validation of \textit{eyeWnder}. Since, however, the heuristic is a reasonable one, and the users have been selected carefully, we assume that their classifications will be more right than wrong. Ads falling under UNKNOWN are ads for which it is impossible to assess precision by the previous means. We handle them separately in Section~\ref{subsec:unknown}. Next, we traverse the evaluation flow-chart of Figure~\ref{fig:evaluationTree} from top to bottom, starting from the right branch.

\noindent \textbf{Ads classified as targeted}: If an ad is classified as targeted by \textit{eyeWnder}, we check to see if the crawler happens to see the same ad. If this happens, then we have a false positive with high probability, since targeted ads should not be encountered by a crawler having empty browsing history.

If the crawler does not see the ad, then we proceed to check \emph{if the ad shares any semantic overlap with the user}, using the methodology described in~\cite{Papaodyssefs_2015}. If it does, then, and only then, we check to see if the content-based heuristic also classifies it as targeted. We do this check in order to identify the set of ads in which \textit{eyeWnder} and the content-based heuristic have a chance to agree.
Such ads can only be \emph{Direct} targeted ads because the content-based heuristic can only detect those (see Section~\ref{subsect:typesoftargeting}).
In our evaluation, we check for semantic overlap using the content-based heuristic which implies that if the ad has semantic overlap with the user, \textit{eyeWnder} and the content based heuristic will agree by default\footnote{It might appear as redundant to check both for semantic overlap and agreement with the content-based heuristic, since we use the latter also for semantic overlap. We have kept both stages for generality, since semantic overlap could be checked with alternative methods than our content-based heuristic.} yielding a likely true positive.

If an ad does not share semantic overlap with the user, we check to see if FigureEight users have tagged it. Notice here that FigureEight users have classified only a subset of the ads that they have seen. If they have not provided a classification, then the ad is added to UNKNOWN. Otherwise, if their classification agrees with \textit{eyeWnder}, we have a likely true positive, else a likely false positive.

\noindent \textbf{Ads classified as non-targeted:} If an ad that \textit{eyeWnder} classifies as non-targeted is seen by the crawler, then this produces a true negative. If the ad has semantic overlap with the user, then, as explained before, the content-based heuristic will classify it as targeted, thereby yielding a likely false negative for \textit{eyeWnder}.
If the ad does not have semantic overlap with the user, we check to see if FigureEight users have classified it. If they have not, it goes to UNKNOWN. If they have classified as targeted, it produces a false negative for \textit{eyeWnder}.
If they have classified it as non-targeted, then we have a likely true negative for \textit{eyeWnder}.

\subsubsection{Dealing with the unknown}
\label{subsec:unknown}

Due to the lack of ground truth, a high rate of classified ads (148 targeted and 4219 non-targeted) have ended up in the two UNKNOWN groups of Figure~\ref{fig:evaluationTree}, for which we cannot evaluate precision using the crawler, the content-based heuristic or FigureEight. In this section, we perform extra analysis to resolve non-targeted UNKNOWNs into likely-TN or likely-FN, and the targeted UNKNOWNs into likely-TP or likely-FP.

\noindent{-- Non-targeted UNKNOWN}: In this case, both count-based and content-based classification have identified these ads as non-targeted.
However, these ads were not manually tagged by FigureEight users. Hence, we select a random set of 200 of these ads and manually inspect them. In particular, we consider the profile of the user receiving the ad, in order to manually determine if the ad is targeting such a profile.

\noindent{-- Targeted UNKNOWN}: This group includes 148 ads that \textit{eyeWnder} classified as targeted, but CB and FigureEight users did not. Hence, they may be either FP or some form of targeting (\eg, indirect) that escapes CB or FigureEight users. A preliminary manual inspection showed that several of them seemed to be retargeted ads~\cite{Bashir_2016}. To verify this, we manually visited the landing page associated to each ad, and afterwards we visited some of the domains where the ad re-appeared according to our dataset.
The experiment was set up for testing the repeatability of the suspected retargeting.
When the experiment led to retargeting, it meant that our initial guess was correct, and we considered the classification to be a likely TP.

For the remaining ads, we evaluated if they could be indirect OBA ads.
To this end, we performed a correlation analysis between the topics of the ad's landing page and the profiles of the users receiving that ad.
If there exists statistically significant correlation between some of the ad's and the user's topics, but these topics are not semantically overlapping, then we interpret it as \emph{likely indirect OBA} ad, and the classification is a likely TP.
Some examples of ads that we identified as indirect OBA are the following:
(1) Male users who exhibit interest in computers, electronics, cars, etc., but receiving ads from a dating website (perhaps a classic indirect targeting of single male users).
(2) Users with interest in computers, electronics, and programming, but receiving ads from KFC, a famous fast food restaurant.
(3) Users with interest in websites related to beauty products, fitness, body care, \etc, receiving ads related to seafood.
Finally (4) users who showed interest in governmental websites, internet services, insurance services, \etc, receiving ads related to real estate and housing.

\subsubsection{Results}

The rates in  Figure~\ref{fig:evaluationTree} along with the results from our analysis of unknown ads reveal that the overall rate of \emph{likely} TPs is 78\%.  From these, 10\%  have been identified by CB or FigureEight users. The rest are  associated to retargeting or indirect OBA.
Our analysis also reveals a TN rate of 87\%. In particular, 27\% are \emph{highly probable} TNs, since these ads have been marked as non-targeted by both, \textit{eyeWnder} and the crawler. The remaining 60\% are \emph{likely} TNs. This  percentage has been derived from our manual inspection of non-targeted UNKNOWN.

Based on these results, we claim that our method meets the high precision requirement defined in Section \ref{sect:background}, which guarantees that, for instance, users willing to report a privacy incident related to an ad have high confidence that the ad is indeed targeted.

We acknowledge that there is still room for improvement, since, as the evaluation shows, \textit{eyeWnder} sometimes fails to detect ads classified as targeted by CB (416) or F8 (91). This is mainly due to two factors. First, as we further discuss in Section~\ref{sect:conclusions}, we have evaluated \textit{eyeWnder} with just a few users, which is close to a worst-case analysis given the crowdsourced nature of the system. Second, \textit{eyeWnder} has been configured to maximize the precision rather than recall for reasons discussed in Section~\ref{sect:background}. We are confident that with a bigger dataset, our existing count-based detection would perform better. We could also come up with more elaborate heuristics to improve performance. Our objective in this paper, however, is to make the point that even with few users, simple count-based detection yields precise classification for both directly and indirectly targeted ads.

\noindent\textbf{Evading detection of targeted ads:} Finally, in this paragraph we discuss the situations where advertisers are intentionally trying to evade the detection of targeted ads. Our system is robust to that since there is a lot of inherent randomness in the current way ads are delivered, yet \textit{eyeWnder} identifies such ads with good accuracy even with few hundreds of users. For an adversary to defeat detection it will have to effectively give up targeting or do it very mildly which would, effectively, go against the very idea of targeting and his business model.

\section{Socio-economic biases}
\label{sect:socioeconomics}

In this section, we leverage the social, economic and demographic information we have from the volunteers and the datasets of Section~\ref{subsect:evaluation_dataset}, to look for potential socio-economic biases of ad targeting.

\begin{table}[t!]
\caption{Logistic regression modeling for targeted ads}
\resizebox{\columnwidth}{!}{%
\begin{tabular}{|c|c|ccclc|}
\hline
\multicolumn{2}{|c|}{Variable}    & OR    & SE    & Z-val  &\multicolumn{1}{c}{P\textgreater{}|z|} & 95\% CI     \\ \hline
\multirow{2}{*}{\rotatebox[origin=c]{30}{Gender}} & female  & 0.255 & 0.407 & -3.356 & 8e-4$^{****}$      & 0.107-0.539 \\
                        & male    & 0.174 & 0.383 & -4.566 & 5e-6$^{****}$      & 0.076-0.348 \\  \hline
\multirow{3}{*}{\rotatebox[origin=c]{30}{Income}} & 30k-60k & 1.446 & 0.145 & 2.538  & 0.0111$^{**}$      & 1.088-1.924 \\
                        & 60k-90k & 1.521 & 0.187 & 2.249  & 0.0245$^{**}$      & 1.052-2.187 \\
                        & 90k-... & 0.525 & 0.343 & -1.878 & 0.0603$^{*}$       & 0.257-0.996 \\ \hline
\multirow{5}{*}{\rotatebox[origin=c]{30}{Age}}    & 20-30   & 1.031 & 0.407 & 0.075  & 0.9404             & 0.488-2.450 \\
                        & 30-40   & 1.428 & 0.405 & 0.880  & 0.3790             & 0.679-3.388 \\
                        & 40-50   & 1.964 & 0.422 & 1.599  & 0.1098             & 0.899-4.788 \\
                        & 50-60   & 0.745 & 0.489 & -0.601 & 0.5475             & 0.291-2.022 \\
                        & 60-70   & 2.654 & 0.477 & 2.044  & 0.0409$^{**}$      & 1.069-7.087 \\ \hline
\end{tabular}
\label{tab:log-reg-model}
}
\begin{center}
\raggedbottom \footnotesize OR: Odds Ratio, SE: Standard Error, CI: Confidence Interval \\
$^{****} p<0.001$, $^{***} p<0.01$, $^{**} p< 0.05$, $^{*} p< 0.1$ 
\end{center}
\end{table}

\subsection{Logistic regression analysis}

The demographic data provided by our volunteers include information such as gender ($G$), age ($A$), income level ($L$), employment status ($E$), \textit{etc.}
We want to investigate these independent, nominal and ordinal factors, and how they associate with the type of advertisement delivered to the participating users.
In practice, the type of ad is the dependent variable ($D$) in our model, receiving a binary status of ``static'' or ``targeted'' advertisement.
Using these input data, we perform a binomial logistic regression on the 4 independent variables to model the 1 dependent variable, in the form: $D \sim G+A+L+E$.

In reality, we tested several configurations for the model, including pairwise interactions between all independent factors, as well as removing incrementally each one of them to test if the model improved its predictive performance.
In fact, in the case of ``employment status'', it was removed from the model as it was deemed non-useful with an $anova$ \emph{likelihood ratio test}, with non-significant impact in the final produced model.
Here, we only report the experimental setup that yielded a logistic regression model with most statistically significant results, in form: $D\sim G+A+L$, considering 0-30k and 1-20 the base levels for income and age, respectively.

\begin{figure}[t!]
\begin{center}
\includegraphics[scale=0.56]{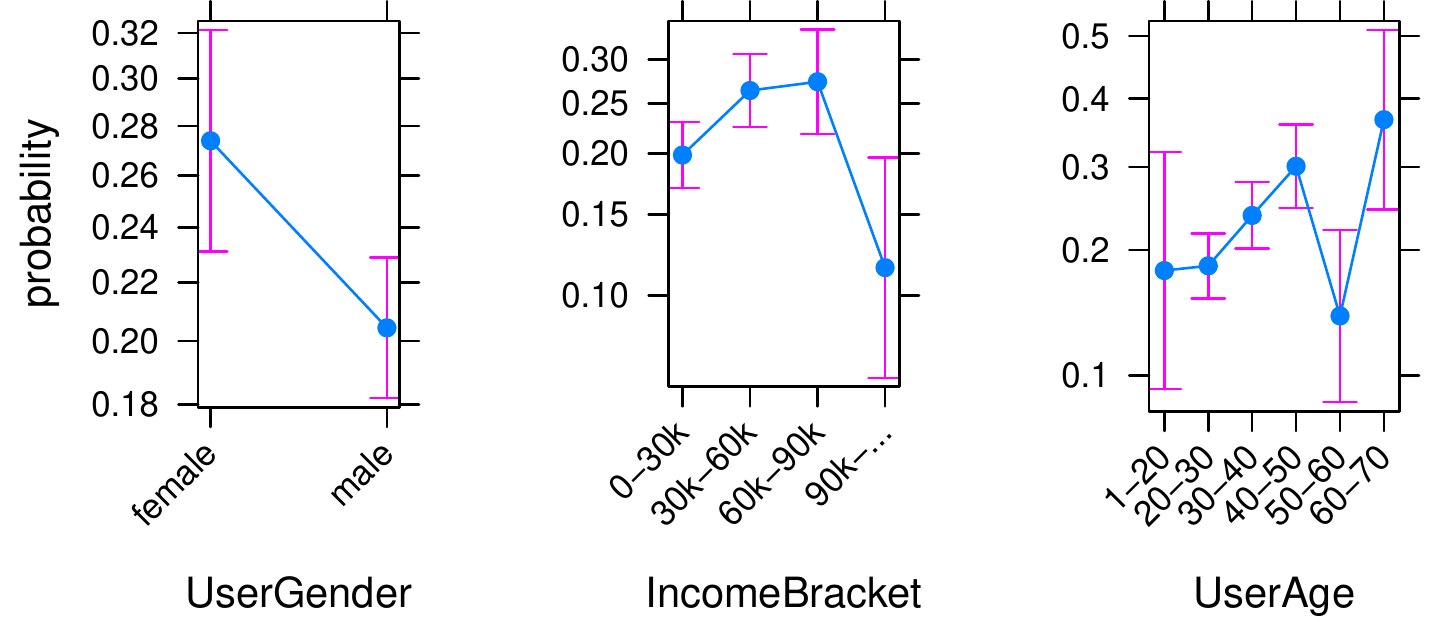}
\caption{Predicted probability for a targeted advertisement to be delivered to a user, vs. three independent variables with statistically significant levels.}
\label{fig:log-reg-effects}
\end{center}
\end{figure}

\subsection{Findings}

The results of the logistic regression are shown in Table~\ref{tab:log-reg-model}.
Also, in Figure~\ref{fig:log-reg-effects} we plot the effects of the three variables and all their corresponding levels, with respect to the expected probability for a user to receive targeted ads.
From these results, we make the following observations:

\noindent -- Gender bias: From our analysis, we find statistically significant effect on the factor of gender.
In particular, women are more likely to be targeted by advertisers than men, and this result is significant at $p<0.001$.

\noindent -- Economic bias: Our results show that as the income level of a user increases from $30k$ to $60k$, and from $60k$ to $90k$ euros, he or she is more likely to be targeted by advertisers, and this is statistically significant at $p<0.05$.
However, when the income level becomes too high (\ie, above $90k$ euros), the expected probability of targeted advertising is reduced.
We conjecture that advertisers profile users to construct economic capacities for each one, and adjust their targeting campaigns accordingly to optimize click-through and sale rates.
Online advertisements targeting very wealthy users may have proven to be less profitable than other income brackets, and thus, such users are less targeted.

\noindent -- Age bias: From the odds ratio scores, it appears that increasingly older users are more probable to be targeted by advertisers.
However, some of the age brackets are less populated by ads and users and thus, our results are not statistically significant, but only demonstrate a consistent trend.

\section{Related Work}
\label{sect:related_work}

\begin{table}[t!]
\footnotesize
\begin{center}
\caption{Comparison of characteristics of main targeted ad detection solutions.}
\resizebox{\columnwidth}{!}{%
\begin{tabular}{|c|c|c|c|c|c|c|c|c|}
\hline
                       & \cite{AdFisher_Datta} & \cite{Adscape:2014} & \cite{AdReveal} & \cite{Carrascosa_2015} & \cite{xray} & \cite{Sunlight} & \cite{MyAdChoices_Jagdish16} & \textit{eyeWnder} \\\hline
Fake Impressions       &\textcolor{red}{$\dagger$} &\textcolor{red}{$\dagger$} &\textcolor{red}{$\dagger$} &\textcolor{red}{$\dagger$} &\textcolor{red}{$\dagger$} &\textcolor{red}{$\dagger$} &\textcolor{red}{$\dagger$} & \\\hline
Click-fraud            & &\textcolor{red}{$\dagger$} & &\textcolor{red}{$\dagger$} & & \textcolor{red}{*} & & \\\hline
Privacy-preserving     & & & & & & & &  \textcolor{green}{\checkmark} \\\hline
Real-users             & & & & & & &  \textcolor{green}{\checkmark} & \textcolor{green}{\checkmark} \\\hline
Personas               &$\bullet$ &$\bullet$ &$\bullet$ & $\bullet$ &$\bullet$ &$\bullet$ & &   \\\hline
Operates in Real-time   & & & & & & & \textcolor{green}{\checkmark} &\textcolor{green}{\checkmark}  \\\hline
High scalability       & & & & & & & \textcolor{green}{\checkmark} &\textcolor{green}{\checkmark}  \\\hline
Operates Offline        & \textcolor{red}{$\dagger$} &\textcolor{red}{$\dagger$} &\textcolor{red}{$\dagger$} & \textcolor{red}{$\dagger$} &\textcolor{red}{$\dagger$} &\textcolor{red}{$\dagger$} & &   \\\hline
Topic-based            &  &$\bullet$ &$\bullet$ &$\bullet$ &  & &$\bullet$ &  \\\hline
Correlation-based      & $\bullet$ & & & & $\bullet$ &$\bullet$ & &  \\\hline
Count-based            & & & & & & & & $\bullet$  \\\hline
\end{tabular}
}
\label{table:comparison_rw}
\end{center}
\begin{center}
\raggedbottom \scriptsize \textcolor{red}{$\dagger$} Negative,  \textcolor{green}{\checkmark} Positive,  $\bullet$ Neutral, \textcolor{red}{*} Unspecified - (Within the context of this work)
\end{center}
\end{table}

Table~\ref{table:comparison_rw} summarizes and compares existing proposals for detecting targeted advertisements. From a methodological view, these solutions can be grouped into \emph{topic-based}~\cite{Adscape:2014, AdReveal, Carrascosa_2015, MyAdChoices_Jagdish16}, and \emph{correlation-based} detection~\cite{AdFisher_Datta, xray, Sunlight}. 

Topic-based solutions perform content-based analysis to extract the relevant topics on a user's browsing history and the ads he receives.
Then, using different heuristics and statistical means, targeted ads are identified as those having topics that share some semantic overlap with the user's browsing history.
Topic-based detection could, in principle, be applied to real users, as we have done for evaluation purposes in Section~\ref{sect:evaluation}.
Existing work, however, has only used it in conjunction with artificially constructed \emph{personas}, \ie, robots that browse the web imitating very specific (single-topic) demographic groups~\cite{Adscape:2014, Carrascosa_2015}, or to emulate real-users offline using click-streams~\cite{AdReveal}. 

The only topic-based solution meant to be used by real users is MyAdchoice~\cite{MyAdChoices_Jagdish16}, which has been implemented in the form of a browser extension. 
This extension is available only under request, and based on the information reported in the paper, it has been only used in a beta-testing phase by few tens of friends and colleagues. 
Independently of the specific pros and cons of individual solutions, topic-based detection presents some common limitations. The most important being that it can only detect direct interest-based targeted advertising. It is unable to detect other forms of targeting based on demographic or geographic parameters, as well as indirect targeting (see Section~\ref{subsect:typesoftargeting} for definitions).

Correlation-based solutions treat the online advertising ecosystem as a blackbox and apply machine learning and statistical methods to detect correlations between the browsing behavior and other characteristics of a user (OS, device type, location, \etc) and the ads he sees. For instance, XRay~\cite{xray} and Sunlight~\cite{Sunlight} create for each persona several \emph{shadow accounts}. 
Each shadow account performs a subset of the actions performed by the original persona. 
By analyzing the common actions performed by shadow accounts receiving the same reaction from the ecosystem (\ie, the same ad), the authors can infer the cause of a targeting event. 
AdFisher~\cite{AdFisher_Datta} uses similar concepts to find discrimination practices, for instance, in the ads shown to men vs. women. 
As with topic-based detection, this technique presents important challenges related to scalability and practical implementation. Moreover, they are not suitable for real-time targeting detection.
With the exception of~\cite{MyAdChoices_Jagdish16}, no previous work has been implemented as a tool for end-users.
Most of them, including~\cite{MyAdChoices_Jagdish16}, rely on content-based analysis, thereby suffering from scalability issues and inability to detect indirect targeting.

Parallel to efforts from the research community, the European Interactive Digital Advertising Alliance (EDAA) has developed YourAdChoices~\cite{adChoice}.
It is a self-regulation program in which companies that deliver targeted ads voluntarily add an icon that, if clicked by a user, offers some form of explanation of why the user received the ad. This technique scales and works in real time.
It only works, however, with companies participating in the program. It also assumes full trust on the reported explanations, something that has been challenged by recent works~\cite{AdFisher_Datta}.
\textit{eyeWnder} offers the opportunity to conduct independent audits, which are useful for end-users, data protection authorities, as well as for the credibility of ad-choices and related self-regulation initiatives. 

\section{Conclusions and future work}
\label{sect:conclusions}

We have showed that a simple count-based heuristic can detect targeted advertising without having to perform complex content-based analysis, \ie without having to understand semantically webpages or received ads. To be able to run such an algorithm, one needs a crowdsourced database of how many users have seen each ad. Such a crowdsourced database can be built efficiently and without jeopardizing user privacy, \ie, without requiring users to report the actual ads they have seen, nor the websites where they encountered them.
Our count-based heuristic and privacy preserving crowdsourced approach have been implemented in a first of its kind distributed system called \textit{eyeWnder}, which allows users to audit any encountered ad impression in real-time to check whether it is targeted or not. We have developed a detailed validation methodology for the difficult problem of assessing the accuracy of an ad detection method using only publicly available information.

Crowdsourcing simplifies the ad detection problem. Any crowdsourcing method, however, has two Achilles' heels: privacy risks and bootstrapping its user-base. We have addressed the first and taken only very preliminary measures for the second. For example, we circulated it among other researchers, and enlisting some users for pay. This has permitted us to conduct a preliminary evaluation of \textit{eyeWnder} and show that the count-based approach is indeed promising.
Our current effort is to scale up our user-base.
To do so, we will use traditional means, \eg, seeking more exposure through media, or getting help from data protection authorities to enlist users.
Scaling up the user-base will help us refine our count-based ad detection method, evaluation, and probably yield many more interesting findings.
This, however, remains a task for future work.

\section*{Acknowledgements}

The research leading to these results has received funding from the European Union's Horizon 2020 Research and Innovation Programme under grant agreements No 653449 (project TYPES), No 786669 (project CONCORDIA), No 786741 (project SMOOTH) and Marie Sklodowska-Curie grant agreement No 690972 (project PROTASIS).
The paper reflects only the authors' views and the Agency and the Commission are not responsible for any use that may be made of the information it contains.

\bibliographystyle{ACM-Reference-Format}
\bibliography{MyCollection}

\end{document}